\documentstyle[prd,aps,floats,psfig]{revtex}
\input epsf
\begin{document}
\preprint{IC-}
\draft

\renewcommand{\topfraction}{0.99}
\renewcommand{\bottomfraction}{0.99}
\twocolumn[\hsize\textwidth\columnwidth\hsize\csname
@twocolumnfalse\endcsname

\title{Supersymmetry breaking and loop corrections at the end of inflation} 

\author{R. Jeannerot$^1$ and J. Lesgourgues$^2$}
\address{~\\$^1$International Centre for Theoretical Physics (ICTP),
P.O. Box 586, Strada Costiera 11, 34014 Trieste, Italy;
~\\$^2$ SISSA/ISAS, Via Beirut 2-4, 34014 Trieste, Italy}

\date{February 7, 2000} 
\maketitle

\begin{abstract}

We show that quantum corrections to the effective potential in
supersymmetric hybrid inflation can be calculated all the way from the
inflationary period - when the Universe is dominated by a false vacuum
energy density - till the fields settle down to the global supersymmetric
minimum of the potential. These are crucial for getting a continuous
description of the evolution of the fields.

\end{abstract}

]

\section{Introduction}

Inflation solves many of the outstanding problems of the standard
cosmology \cite{Lindebook}. Among others, it provides a mechanism for
 the generation of primordial cosmological perturbations, which are
responsible for the observed temperature anisotropies in the cosmic
microwave background (CMB), and the large-scale structure (LSS) in our
Universe. Future experiments, such as
MAP\footnote{http://map.gsfc.nasa.gov} and 
Planck\footnote{http://astro.estec.esa.nl/SA-general/Projects/Planck}
(resp. 2dF\footnote{http://meteor.anu.edu.au/~colless/2dF} and
SDSS\footnote{http://www.astro.princeton.edu/BBOOK}), will measure
with great precision the power spectrum of the CMB (resp. LSS), and
draw sharp constraints on the potential of the inflaton, the scalar
field driving inflation.

Successful implementation of the inflationary picture requires a long
enough era of accelerated expansion on one hand, and a correct order
of magnitude for primordial perturbations on the other hand. In the
most simple versions of single-field inflation, the corresponding
constraints on the inflaton potential are unrealistic from a particle
physics point of view. Indeed, coupling parameters must be fine-tuned
to very small values, while the inflaton must be of the order of the
Planck mass during inflation. These problems can be avoided in
the so-called hybrid models \cite{Lindehybrid} (see also Ref.
\cite{Cop}), in which the inflaton couples with some other(s) scalar
field(s). 
Hybrid inflation arises naturally in supersymmetric
theories\footnote{For a review on inflation in supersymmetric theories
see Ref. \cite{LythRiotto}.} \cite{Cop,DvaSha,stewart,Dterm}. Supersymmetry
provides the flatness of the scalar potential required for
inflation. The inflaton field, which is usually a scalar singlet,
couples to Higgs superfield(s) charged under some gauge group G. At
the end of inflation, the Higgs fields acquire a non-vanishing vacuum
expectation value (VEV); G is spontaneously broken. Such scenarios
arise naturally in supersymmetric grand unified theories (SUSY GUT)
\cite{DvaSha,Rinf}. The non-zero vacuum energy density during
inflation can either be due to the VEV of a F-term \cite{Cop,DvaSha}
or from that of a D-term \cite{Cop,stewart,Dterm}.

Models of either types share the following features. The scalar
potential has two minima; one local minimum for values of the inflaton
field $|S|$ greater than some critical value $s_c$ with the Higgs
fields at zero, and one global supersymmetric minimum at $S=0$ with
non-zero Higgs VEVs. The fields are usually assumed to have chaotic
initial conditions \cite{Lindechaotic}, with an initial value $|S| \gg
s_c$ for the inflaton. The Higgs fields rapidly settle down to the
local minimum of the potential\footnote{The problem of initial
conditions in inflation is actually not trivial. We shall not discuss
this problem here. See Refs. \cite{Tetradis} for possible solutions.};
then, the universe is dominated by a non-vanishing vacuum energy
density, and supersymmetry is broken. This in turn leads to quantum
corrections to the potential which lift its complete flatness
\cite{DvaSha,Dvali}. The slow-roll conditions are satisfied and
inflation takes place until $|S|=s_c$ or slighlty before, depending on
the model. When $|S|$ falls below $s_c$, the Higgs fields start to
acquire non-vanishing VEVs. All fields then oscillate until they
stabilise at the global supersymmetric minimum. These oscillations are
crucial for understanding the process of reheating and particle
production in the early universe.

The important point is that, as long as the fields are not settled
down at the global minimum, supersymmetry remains broken. When the
fields oscillate, the system only goes ponctually through
supersymmetric configurations. The breaking of supersymmetry is best
seen by looking at the mass spectrum; the bosonic and fermionic masses
are non degenerate. This has important consequences. It implies that
loop corrections to the effective potential are non-zero not only
during inflation, but also during all the oscillatory regime. The
corrections are crucial for getting a continuous description of the
evolution of the fields. They will be useful for the simulation of
preheating, for the calculation of the number density of cosmic
strings, for the study of leptogenesis at the end of inflation
\cite{Sha,Rlept}, and for the derivation of the primordial spectrum
during intermediate stages in supersymmetric multiple inflationary
models \cite{TetSak,Lesgourgues99}.

In this Letter, we calculate the one-loop corrections to the potential
along the inflaton direction. They are the most important and affect
the dynamics of the inflaton field, which in turn affects the
dynamics of the Higgs fields. They can be calculated by
applying the Coleman-Weinberg formula \cite{ColWey}:
\begin{equation}
\Delta V = \frac{1}{ 64 \pi^2} \sum_i (-1)^F m_i^4 \ln
(m_i^2/\Lambda^2), \label{eq:CW}
\end{equation}
where $(-1)^F$ shows that bosons and fermions
make opposite contributions; it is $+1$ for the bosonic degrees of
freedom and $-1$ for the fermionic ones. The sum runs over each degree
of freedom $i$ with mass $m_i$ and $\Lambda$ is a renormalization
scale.  We thus determine the particle
spectrum for each value of the inflaton field $|S|$ and values of the
other fields which minimize the potential for this $|S|$. We consider
the standard models of F and D-term inflation. The particle spectrum
is found to be very rich and interesting. During inflation, the
non-zero quantum corrections are due to a boson-fermion mass splitting
in the Higgs sector. When $|S|$ falls below $s_c$, since the Higgs
VEVs are non zero, there is also a mass splitting between gauge
and gaugino fields.

\section{F-term inflation}

The simplest superpotential which leads to F-term inflation is given
by $W = \alpha S \overline{\Phi} \Phi - \mu^2 S$, where $S$ is a
scalar singlet and ($\Phi$, $\overline{\Phi}$) are Higgs superfields
in complex conjugate representations of some gauge group G
\cite{Cop,DvaSha}. $\alpha$ and $\mu$ are two constants which are
taken to be positive and ${\mu \over \sqrt{\alpha}}$ sets the G
symmetry breaking scale. This superpotential is consistent with a
continuous R symmetry under which the fields transform as $S
\rightarrow e^{i\gamma} S$, $\Phi \rightarrow e^{i\gamma} \Phi$,
$\overline{\Phi} \rightarrow e^{-i\gamma} \overline{\Phi}$ and $W
\rightarrow e^{i\gamma} W$. It is is often used in SUSY GUT model
building.  The scalar potential reads:
\begin{eqnarray}
V &=& \alpha^2 |S|^2 (|\overline{\Phi}|^2 + |\Phi |^2) + |\alpha
\overline{\Phi} \Phi -\mu^2|^2 \nonumber \\
&+& {g^2\over 2} (|\overline{\Phi}|^2 -
|\Phi |^2)^2,  \label{eq:pot}
\end{eqnarray}
where we have kept the same notation for the superfields and their
bosonic components. There is a flat direction with degenerate local
minima $|S| \equiv s \geq s_c = {\mu \over \sqrt{\alpha}}$,
$\overline{\Phi} = \Phi = 0$, for which $V=\mu^4$, and a global
supersymmetric minimum at $S=0$, $|\Phi | = |\overline{\Phi}| = {\mu
\over \sqrt{\alpha}}$, $\arg (\Phi ) + \arg ( \overline{\Phi}) =0$,
in which the G symmetry is spontaneously broken.

We assume that the problem of initial conditions has been solved, and
we investigate the behavior of the system already settled in the local
minimum of the potential. The universe is dominated by the vacuum
energy density $V=\mu^4$ and supersymmetry is broken.  The bosonic and
fermionic masses are thus non degenerate. The mass splitting happens
in the $\Phi$ and $\overline{\Phi}$ sector. Explicitly, there are two
complex scalars with masses squared $m_1^2 = \alpha^2 s^2 + \mu^2
\alpha$ and $m_2^2 = \alpha^2 s^2 - \mu^2 \alpha$ (the mass
eigenstates are linear combinaisons of the $\Phi$ and
$\overline{\Phi}$ fields), and two Weyl fermions with masses $m^2 =
\alpha^2 s^2$. This spectrum gives rise to quantum corrections to the
effective potential which can be calculated from Eq.(\ref{eq:CW}) and
lift the the complete flatness of the $s$ direction. When $s \gg s_c$,
they have a well-know asymptotic form \cite{DvaSha}:
\begin{equation}
\Delta V = {\alpha^2 \mu^4 \over 16 \pi^2} \left ( \ln {\alpha^2 s^2 \over
\Lambda^2} + {3\over 2} \right ). \label{eq:cor}  
\end{equation}
Therefore, the $S$ field can roll down the potential. The slow roll
conditions are satisfied and inflation takes place. When $s$ falls
below $s_c$, $\Phi$ and $\overline{\Phi}$ are destabilized, and all
fields start to oscillate.

During inflation, the Higgs fields $\Phi$ and $\overline{\Phi}$ have
zero VEVs and since the inflaton $S$ is assumed to be a gauge singlet,
gauge and gauginos have zero masses; the only contribution to $\Delta
V$ comes from the mass splitting in the $\Phi$ and $\overline{\Phi}$
sector. Now when $s$ falls below $s_c$, the VEVs of $\Phi$ and
$\overline{\Phi}$ start to be non-zero. Thus the corresponding gauge
and gaugino fields, as well as the $S$ field, also acquire non-zero
mass. The mass splitting then happens both in the Higgs and in the
gauge sectors.

From now on, we shall assume that the VEVs of $\Phi$ and
$\overline{\Phi}$ only break a U(1) gauge symmetry, and that the
representation of $\Phi$ is complex one-dimensional. For arbitrary
$n$-dimensional complex conjugate representations which break a gauge
group G down to a subgroup H of G, when the Higgs VEVs are non-zero,
there are $k= {\rm dim}(G) - {\rm dim}(H)$ massive gauge fields and
$4n - k + 2$ massive real scalar fields. For any value of $s$, the
potential is minimised along the D-flat direction $|\overline{\Phi}| =
|\Phi | = \hat{\phi}$, and for $\arg (\Phi ) = - \arg
(\overline{\Phi}) = \theta$. Therefore, we expand the fields as
follows:
\begin{equation}
\Phi = \hat{\phi} e^{i\theta} + \phi_1, \qquad
\overline{\Phi} = \hat{\phi} e^{-i\theta} + \phi_2, 
\end{equation}
where $\phi_1$ and $\phi_2$ are complex fields which represent the
quantum fluctuations of the $\Phi$ and $\overline{\Phi}$ fields.
$\hat{\phi} = 0$ for $s \geq s_c$ and $\hat{\phi} = \sqrt{{\mu^2 -
\alpha s^2 \over \alpha}}$ for $s \leq s_c$.  When $s$ falls below
$s_c$, we find the following spectrum.  There is a complex scalar
field with squared mass $m_S^2 = 2 \alpha^2 \hat{\phi}^2$. The Higgs
mechanism gives rise to three real scalars with masses squared
$m_{1}^2 = 2 \alpha^2 \hat{\phi}^2$, $m_{2}^2 = 2 \alpha \mu^2$ and
$m_{3}^2 = 2 \alpha^2 s^2 + 4 g^2 \hat{\phi}^2$. The corresponding
Higgs mass eigenstates are $\Re (\phi_1) + \Re (\phi_2)$, $\Im
(\phi_1) + \Im (\phi_2)$ and $ \Re (\phi_1) - \Re (\phi_2)$. The field
$\Im (\phi_1) - \Im (\phi_2)$ is absorbed by the gauge field which is
now massive with mass squared $m_A^2 = 4 g^2 \hat{\phi}^2$.

The fermionic spectrum can be derived
from the following parts of the Lagrangian:
\begin{eqnarray}
{\cal L}_Y &=& \alpha (S \psi_1 \psi_2 + \Phi \psi_S \psi_2 +
\overline{\Phi} \psi_1 \psi_S), \nonumber \\
{\cal L}_g &=& - i \sqrt{2} g (\tilde{\Lambda} \psi_2 \overline{\Phi}^* -
\tilde{\Lambda} \psi_1 \Phi^*) + {\mathrm h.c.}, 
\end{eqnarray}
where $\psi_1$, $\psi_2$ and $\psi_S$ are the fermionic components of
the Higgs and inflaton
superfields. $\tilde{\Lambda}$ is the gaugino. After diagonalizing the
fermion mass matrix, we find that there are four Weyl fermions with
masses:
\begin{eqnarray}
m^2_{\psi_1^\pm} &=& 2 \alpha^2 \hat{\phi}^2 + {\alpha^2 s^2 \over 2}
\pm {1\over 2} \alpha^2 s \sqrt{8 \hat{\phi}^2 + s^2},
\nonumber \\
m^2_{\psi_2^\pm} &=&  4 g^2 \hat{\phi}^2 + {\alpha^2 s^2 \over 2} \pm {1\over
2} \alpha s \sqrt{16 g^2 \hat{\phi}^2 + \alpha^2 s^2}.
\end{eqnarray}
The mass eigenstates are, respectively, linear combinations of
the higgsinos and the inflatino (the fermionic component of the inflaton
superfield), and linear combinations of the higgsinos and the gaugino. We
summarize our results in Tables \ref{tab:TF1} and \ref{tab:TF2}.  
The one-loop corrected effective potential is $V + \Delta V(s)$, where
$\Delta V(s)$ is given by Eq. (\ref{eq:CW}).  We check that the
supertrace ${\mathrm Str} M^2 \equiv \sum_i (-1)^F m^2_i$ vanishes at
all times \cite{Ferrara}, and that the corrections are
continuous at $s=s_c$. The exact effective potential should be a
smooth function of $s$, and independent of $\Lambda$. In the one-loop
approximation, $\Lambda$ must be chosen so that the contribution of
higher order terms can be neglected. This is generally achieved with
$\Lambda^2 \sim \alpha^2 s_c^2 = \mu^2 \alpha$ \cite{Lyth}. Here, by
imposing the continuity of the potential derivative at $s=s_c$, we find:
\begin{equation}
\Lambda^2 = e^\epsilon \alpha^2 s_c^2, \qquad \epsilon \equiv \frac{1}{2}
- \frac{g^2 \ln 2}{\alpha^2+g^2}.
\end{equation} 
When $g \leq \alpha$, the shape of $\Delta V(s)$ is given in
Fig. \ref{fig}, and $|\Delta V(s)| \ll V$. When $g > \alpha$, a higher
mass scale $g^2 \hat{\phi}^2 > \alpha^2 s_c^2$ appears after symmetry
breaking; thus, the above choice of $\Lambda^2$ is not valid anymore, and
we find that the one-loop corrections exceed the tree-level potential;
so, in the one-loop approximation, it is impossible to find an
expression for $\Delta V (s)$ valid around $s_c$.
\begin{table}\begin{center}
\begin{tabular}{c c c}
Field & d.o.f. & squared mass  \\
\hline 
$ {\phi_1^* + \phi_2 \over \sqrt{2}}$  & 2 & $\alpha^2 s^2 + \mu^2
\alpha$  \\ \hline 
$ {- \phi_1^* + \phi_2 \over \sqrt{2}}$  & 2 & $\alpha^2 s^2 - \mu^2
\alpha$  \\ \hline
2 Weyl fermions & $2 \times 2$ & $\alpha^2 s^2$ \\
\end{tabular}\end{center}
\caption{Particle spectrum during F-term inflation
and numbers of degrees of freedom (d.o.f.), when $s = s \geq
s_c$.} \label{tab:TF1} 
\end{table}
\begin{table}[h] \begin{center}
\begin{tabular}{c c c}
Field & d.o.f. & squared mass  \\
\hline 
S & 2 & $2 \alpha^2 \hat{\phi}^2 $ \\ \hline
$\Re (\phi_1) + \Re (\phi_2)$ & 1 &$2 \alpha^2 \hat{\phi}^2 $ \\ \hline
$\Im (\phi_1) + \Im (\phi_2)$ & 1 & $2 \alpha \mu^2$ \\ \hline
 $\Re (\phi_1) - \Re (\phi_2)$ & 1 & $2 \alpha^2 s^2 +
4 g^2 \hat{\phi}^2$  \\ \hline
$A$ & 3 & $4 g^2 \hat{\phi}^2$ \\ \hline
2 Weyl fermions & $2 \times 2$ & $2 \alpha^2 \hat{\phi}^2 + {\alpha^2
s^2 \over 2} \pm { \alpha^2 s \over 2} \sqrt{8 \hat{\phi}^2 + s^2}$
\\ \hline
2 Weyl fermions & $2 \times 2$ & $4 g^2 \hat{\phi}^2 + {\alpha^2 s^2
\over 2} \pm { \alpha s \over 2} \sqrt{16 g^2 \hat{\phi}^2 + \alpha^2 
s^2}$ \\
\end{tabular}\end{center}
\caption{Particle spectrum along the inflaton direction at the end of
F-term inflation, when $\hat{\phi} = \sqrt{{\mu^2 - \alpha s^2} \over
\alpha}$ and $s = s \leq s_c$.} \label{tab:TF2} 
\end{table}

\section{D-term inflation}

A generic toy-model of D-term inflation, proposed in
Refs. \cite{Dterm}, involves three complex fields: a gauge singlet
$S$, and two fields $\Phi_+$ and $\Phi_-$ with charges $+1$ and $-1$
under a $U(1)$ gauge symmetry.  The superpotential is
$W=\lambda S \Phi_+ \Phi_-$. This choice can be justified by a set of
continuous R-symmetries or discrete symmetries. The scalar potential
is:
\begin{eqnarray}
V &=& \lambda^2 |S|^2 (| \Phi_+ |^2  + |\Phi_-|^2) 
+  \lambda^2 | \Phi_+ |^2 |\Phi_-|^2 \nonumber \\
& & + \frac{g^2}{2} ( | \Phi_+ |^2  - |\Phi_-|^2 + \xi )^2,
\end{eqnarray}
where $\xi$ is a Fayet-Illiopoulos term. We suppose that $\xi>0$ (if
it is not the case, the role of $\Phi_+$ and $\Phi_-$ are just
inverted). There is a true supersymmetric vacuum at $S=\Phi_+=0$,
$|\Phi_-|=\sqrt{\xi}$, and a valley of local minima $|S| \equiv s >
s_c = g \sqrt{\xi} / \lambda$, $\Phi_+=\Phi_-=0$, in which the
tree-level potential is flat, $V=g^2 \xi^2 /2$, and the dynamics of
$s$ is only driven by quantum corrections. In the false vacuum,
($\Phi_+$, $\Phi_-$) have squared masses $\lambda^2 s^2 \pm g^2 \xi$,
while their fermionic superpartners $(\psi_+,\psi_-)$ combine to form
a Dirac spinor with mass $\lambda s$ (see Table \ref{tab:TF3}). The
loop corrections are given by Eq. (\ref{eq:CW}), and when $s \gg s_c$
they reduce to:
\begin{equation} \label{deltaV}
\Delta V = \frac{g^4 \xi^2}{16 \pi^2} \left( \ln
\frac{\lambda^2 s^2}{\Lambda^2} + \frac{3}{2} \right).
\end{equation}

\begin{figure}[htb]
\epsfxsize=8cm
$$
\epsfbox{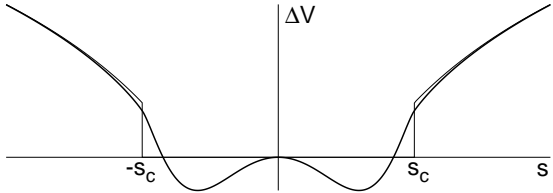}
$$
\caption[]{Shape of the one-loop corrections, for F-term inflation
with $g \leq \alpha$, and D-term inflation. The thin line shows the
approximation of Eqs. (\ref{eq:cor},\ref{deltaV}) for $|S| \geq s_c$.}
\label{fig}
\end{figure}

When $s$ falls below $s_c$, $\Phi_-$ acquires a non-vanishing VEV
$\hat{\phi} e^{i \theta}$, with $\hat{\phi} = \sqrt{ \xi -
\frac{\lambda^2}{g^2} s^2}$, which breaks the $U(1)$ gauge
symmetry. The mass splitting then also happens in the gauge
sector. Without loss of generality, we assume that $\theta = 0$, and
expand the Higgs field as $\Phi_- = \hat{\phi} + \phi_1$. The real
field $\sqrt{2} \Re (\phi_1)$ has a squared mass $2 g^2 \hat{\phi}^2$,
while the Goldstone boson $\sqrt{2} \Im (\phi_1)$ is eaten up by the
gauge boson, which becomes massive with $m^2_A=2 g^2
\hat{\phi}^2$. The masses for $S$ and $\Phi_+$ are given in Table
\ref{tab:TF4}.  The fermionic spectrum can be derived from the
following parts of the supersymmetric Lagrangian:
\begin{eqnarray}
{\cal L}_Y &=& \lambda (S \psi_+ \psi_- + \Phi_+ \psi_S \psi_- +
\Phi_- \psi_S \psi_+ ), \nonumber \\ 
{\cal L}_g &=& - i \sqrt{2} g
(\tilde{\Lambda} \psi_- \Phi_-^* + \tilde{\Lambda} \psi_+ \Phi_+^*) 
+ {\mathrm h.c.}
\end{eqnarray}
\begin{table} 
\begin{tabular}{c c c}
Field & d.o.f. & squared mass \\
\hline
$\Phi_+$ & 2 & $\lambda^2 s^2 + g^2 \xi $   \\
\hline
$\Phi_-$ & 2 & $\lambda^2 s^2 - g^2 \xi $   \\
\hline
Dirac fermion & 4 & $\lambda^2 s^2$          
\end{tabular}
\caption{Particle spectrum during D-term inflation, when
$s \geq s_c$.} \label{tab:TF3}
\end{table}
\begin{table} 
\begin{tabular}{c c c }
Field & d.o.f. & squared mass   \\ \hline
$S$      & 2 & $\lambda^2 \hat{\phi}^2 $  \\ \hline
$\Phi_+$ & 2 & $\lambda^2 \hat{\phi}^2 +
2 \lambda^2 s^2$ \\ \hline
$\sqrt{2} \Re(\phi_1)$ 
& 1 &  $2 g^2 \hat{\phi}^2$ \\ \hline
$A$ & 3 & $2 g^2 \hat{\phi}^2$ \\ \hline
Dirac fermion & 4 & 
$(\frac{\lambda^2}{2} + g^2) \hat{\phi}^2 
+ \frac{\lambda^2}{2} s^2 + \sqrt{\Delta}$ \\ \hline
Dirac fermion & 4 & 
$(\frac{\lambda^2}{2} + g^2) \hat{\phi}^2 
+ \frac{\lambda^2}{2} s^2 - \sqrt{\Delta}$ \\
\end{tabular}
\caption{Particle spectrum along the inflaton direction at the
end of D-term inflation, with $s \leq s_c$ and
$\hat{\phi} = (\xi - (\lambda / g )^2 s^2 )^{1/2}$.
For the fermion masses, we defined $\Delta
\equiv (\frac{1}{2} \lambda^2 
(\hat{\phi}^2 + S^2) 
+ g^2 \hat{\phi}^2 )^2 - 2
\lambda^2 g^2 \hat{\phi}^4 $.} \label{tab:TF4} 
\end{table}
We find that the gaugino, the inflatino and the higgsinos combine to
form two Dirac spinors with masses given in Table \ref{tab:TF4}.
The one-loop effective potential reads:
\begin{eqnarray}
V &=& \lambda^2 s^2 | \Phi_- |^2 + \frac{g^2}{2} (|\Phi_- |^2-\xi)^2
+ \Delta V (s),
\end{eqnarray}
where $\Delta V(s)$ is given by Eq. (\ref{eq:CW}). The supertrace
vanishes at any time. The potential is continuous at $s=s_c$, and 
so is its derivative if we take: 
\begin{equation}
\Lambda^2 = e^\epsilon \lambda^2 s_c^2, \qquad \epsilon \equiv
\frac{1}{2}+ \frac{\ln 2}{3} (1-\frac{\lambda^2}{g^2}).
\end{equation} 
Then, for any choice of $\lambda$ and $g$, the corrections
are small with respect to the tree-level potential, and have the
shape given in Fig. \ref{fig}. 
Let us now comment on the importance of the one-loop corrections to
the first and second derivative at the end of inflation. Obviously,
they are dominant at the very begining of the symmetry breaking.  At
this time, it is important to know exactly $\partial \Delta V /
\partial s$, in order to characterize the emergence of an effective
Higgs background (from the coarse-graining of large-scale quantum
fluctuations). After this stage, when the Higgs start(s) to grow, the
loop corrections will affect $\partial V / \partial s$ and $\partial^2
V / \partial s^2$ only by a few percent\footnote{For instance, in the
D-term case, when the inflaton stabilizes around zero, its oscillation
frequency is usually calculated from the tree-level effective mass: $
\frac{\partial^2 V}{\partial s^2} = 2 \lambda^2 |\Phi_-|^2= 2
\lambda^2 \xi.  $ As can be seen form Fig.1., the loop corrections
will lower this value.  We find:
\begin{eqnarray}
\frac{\partial^2 \Delta V}{\partial s^2} &=& - \frac{g^2 \lambda^2 \xi}{\pi^2}
f ( \frac{\lambda^2}{2g^2} ), \nonumber \\
f(x) &\equiv& \frac{\ln 2}{3} (1+x) - \frac{x \ln x}{2 (1-x)}, 
\qquad f(1) \simeq 1. \nonumber
\end{eqnarray}
So, with $\lambda=g=1$, the effective mass is lowered by 4 \%.
A similar order of magnitude is found in the case of F-term inflation, 
for which
$\frac{\partial^2 V}{\partial s^2} = 4 \alpha \mu^2$:
$$
\frac{\partial^2 \Delta V}{\partial s^2} =
- \frac{ g^2 \alpha \mu^2}{\pi^2} \tilde{f}(\frac{\alpha^2}{g^2}),
$$
and $\tilde{f}(x)$ is a complicated function, of order one when
$1 < x <100$. When $\alpha=2g=1$, the effective mass is lowered by 2 \%.}. 

\section{Conclusion}

In this paper, we have discussed the problem of supersymmetry breaking
during and at the end of supersymmetric hybrid inflation, when the
inflation scale is generated by GUT physics. We did not consider
inflation models at intermediate or low energy scales
\cite{lowinfl,king}. Neither did we include supergravity
corrections. When global supersymmetry is replaced by supergravity,
the F-flat directions can only be preserved for specific K\"ahler
potentials \cite{Cop,stewart}; this does not apply to the case of
D-term inflation, for which supergravity corrections are small for all
values of the fields below the Planck mass \cite{Cop,stewart}.  In
this framework, we calculated the one-loop corrections which modify
the effective potential in the flat (inflaton) direction. We would
like to point out that the classical trajectories of the Higgs fields
do not necessarily coincide with their valley of local
minima. However, for a wide range of parameters, the trajectory
remains very close to this valley.  Also, in more realistic cases, the
Higgs fields will belong to non-trivial representations of G and their
VEVs will break a non-abelian gauge group
\cite{DvaliPLB,Jeannerot53,DvaRio,Masiero}. They may also couple to
fermions such as right-handed neutrinos \cite{Sha,Jeannerot53}. Hence,
generally, one would find a much richer spectrum which would increase
the corrections in a model-dependent way. Our results do not apply
only to the end of inflation. They could be used in other models
for which inflation takes place in local minima that
spontaneously break both the gauge symmetry and supersymmetry. Also,
in supersymmetric multiple inflationary models, part of the gauge
symmetry breaks between the different stages of inflation, and at each
step, the calculation of loop corrections can be performed as
discussed in this paper.

\section*{Acknowledgements}

The authors would like to thank D. Demir, G. Dvali and
G. Senjanovi\'{c} for very useful discussions. J. L. is supported by
the TMR grant ERBFMRXCT960090.


\begin{thebibliography} {99}
\bibitem{Lindebook} {\em Particle physics and inflationary cosmology},
A.D. Linde, Harwood (1990), (Contemporary concepts in physics, 5).
\bibitem{Lindehybrid}  A.D. Linde, Phys. Lett. B {\bf 259}, 38 (1991);
Phys. Rev. D {\bf 49}, 748 (1994).
\bibitem{Cop} E.J. Copeland, A.R. Liddle, D.H. Lyth, E.D. Stewart and
D. Wands, Phys. Rev. D {\bf 49}, 6410 (1994).
\bibitem{LythRiotto} D.H. Lyth and A. Riotto,
Phys. Rept. {\bf 314}, 1 (1999).
\bibitem{DvaSha}  G. Dvali, Q. Shafi and R. Schaefer,
Phys. Rev. Lett. {\bf 73}, 1886 (1994). 
\bibitem{stewart} E. D. Stewart, Phys. Rev. D {\bf
51}, 6847 (1995). 
\bibitem{Dterm} E. Halyo, Phys. Lett. B {\bf 387}, 43 (1996);
P. Bin\'etruy and G. Dvali, Phys. Lett. B {\bf 388}, 241 (1996);
E. Halyo, Phys.Lett. B {\bf 454}, 223 (1999).
\bibitem{Rinf} R. Jeannerot, Phys. Rev. D {\bf 56}, 6205 (1997).
\bibitem{Lindechaotic} A.D. Linde, Phys. Lett. B {\bf 129}, 177 (1983).
\bibitem{Tetradis} C. Panagiotakopoulos and N. Tetradis, Phys. Rev. D
{\bf 59}, 3502 (1999); G. Lazarides and N. D. Vlachos, Phys. Rev. D
{\bf 56}, 4562 (1997); Z. Berezhiani, D. Comelli and N. Tetradis,
Phys. Lett. B {\bf 431}, 286 (1998).
\bibitem{Dvali} G. Dvali, Phys. Lett. B {\bf 355}, 78 (1995); Phys. Lett. 
B {\bf 387}, 471 (1996).
\bibitem{Sha} G. Lazarides, Q. Shafi and N.D. Vlachos,
Phys. Lett. B {\bf 427}, 53 (1998);  G. Lazarides, to appear in
Springer Tracts in Modern Physics: Symmetries in Physics and
Conservation Laws. 
\bibitem{Rlept} R. Jeannerot, Phys. Rev. Lett. {\bf 77}, 3292 (1996).
\bibitem{TetSak}  M. Sakellariadou and N. Tetradis, preprint hep-ph/9806461. 
\bibitem{Lesgourgues99} J. Lesgourgues, Phys. Lett. B {\bf 452}, 15 (1999);
preprint hep-ph/9911447.
\bibitem{ColWey} S. Coleman and S. Weinberg, Phys. Rev. D {\bf 7},
1888 (1973).
\bibitem{Ferrara} S.~Ferrara, L.~Girardello and F.~Palumbo,
Phys. Rev. D {\bf 20}, 403 (1979).
\bibitem{Lyth} D. Lyth, hep-ph/9904371.
\bibitem{lowinfl} L. Randall, M. Soljacic and A.H. Guth,
Nucl. Phys. B {\bf 472}, 377 (1996).
\bibitem{king}M. Bastero-Gil and S.F. King, Phys. Lett. B {\bf 423}, 27
(1998). 
\bibitem{Lazsugra} A. Linde and A. Riotto, Phys. Rev. D {\bf 56}, 1841
(1997); G. Lazarides and N. Tetradis, Phys. Rev. D {\bf 58}, 3502 (1998). 
\bibitem{DvaliPLB} G. Dvali, Phys. Lett. B {\bf 387}, 471 (1996).
\bibitem{Jeannerot53}R. Jeannerot, Phys. Rev. D {\bf 53}, 5426 (1996).
\bibitem{DvaRio} G. Dvali and A. Riotto, Phys. Lett. B {\bf
417}, 20 (1998). 
\bibitem{Masiero} L. Covi, G. Mangano, A. Masiero and G. Miele,
Phys. Lett. B {\bf 424}, 253 (1998). 
\end{thebibliography}
\end{document}